\documentclass[aps,pra,reprint,twocolumn,superscriptaddress,floatfix,nofootinbib,longbibliography]{revtex4-1}
\usepackage{graphicx,amsmath,amsfonts,amssymb,amsthm,xr}
\usepackage{epsfig,amsmath,amssymb,color,dsfont,upgreek,physics}
\usepackage{mathrsfs}
\usepackage{mathtools}
\usepackage{bbold}
\usepackage{comment}
\usepackage{float}

\usepackage[bookmarks=true,colorlinks,linkcolor=OrangeRed,urlcolor=NavyBlue,citecolor=RoyalBlue]{hyperref}
\usepackage[dvipsnames]{xcolor}
\usepackage{orcidlink}

\definecolor{mygold}{rgb}{0.93,0.69,0.13}

\definecolor{mypurple}{rgb}{0.49,0.18,0.56}

\definecolor{mygreen}{rgb}{0,0.5,0}

\usepackage{braket}

\newcommand{\be}{\begin{equation}}
\newcommand{\ee}{\end{equation}}

\newcommand{\eea}{\end{eqnarray}}

\newcommand{\ipr}{{i^{\prime}}}

\begin{document}

\title{Protecting Hilbert space fragmentation through quantum Zeno dynamics}

\author{Pranay Patil${}^{\orcidlink{0000-0002-4554-6539}}$}
\email{patil@pks.mpg.de}
\affiliation{Max Planck Institute for the Physics of Complex Systems, N\"othnitzer Stra\ss e 38, 01187 Dresden, Germany}

\author{Ayushi Singhania${}^{\orcidlink{0000-0001-9617-3026}}$}
\email{singhania@ifw-dresden.de}
\affiliation{Institute for Theoretical Solid State Physics, IFW Dresden, Helmholtzstra\ss e 20, 01069 Dresden, Germany}

\author{Jad C.~Halimeh${}^{\orcidlink{0000-0002-0659-7990}}$}
\email{jad.halimeh@physik.lmu.de}
\affiliation{Department of Physics and Arnold Sommerfeld Center for Theoretical Physics (ASC), Ludwig-Maximilians-Universit\"at M\"unchen, Theresienstra\ss e 37, D-80333 M\"unchen, Germany}
\affiliation{Munich Center for Quantum Science and Technology (MCQST), Schellingstra\ss e 4, D-80799 M\"unchen, Germany}

\begin{abstract}
Hilbert space fragmentation is an intriguing paradigm of ergodicity breaking in interacting quantum many-body systems with applications to quantum information technology, but it is usually adversely compromised in the presence of perturbations. In this work, we demonstrate the protection of constrained dynamics arising due to a combination of mirror symmetry and Hilbert space fragmentation by employing the concept of quantum Zeno dynamics. We focus on an Ising spin ladder with carefully chosen quantum fluctuations, which in the ideal case guarantee a perfect disentanglement under Hamiltonian dynamics for a large class of initial conditions. This is known to be a consequence of the interplay of Hilbert space fragmentation with a mirror symmetry, and we show numerically the effect of breaking the latter. To evince the power of this perfect disentanglement, we study the effect of generic perturbations around the fine-tuned model, and show that we can protect against the undesirable growth of entanglement entropy by using a local Ising interaction on the rungs of the ladder. This allows us to suppress the entanglement entropy to an \textit{arbitrarily} small value for an \textit{arbitrarily} long time by controlling the strength of the rung interaction. Our work demonstrates the experimentally feasible viability of quantum Zeno dynamics in the protection of quantum information against thermalization.
\end{abstract}

\maketitle

\section{Introduction}
One of the most intriguing questions in quantum many-body physics is the nature of thermalization, or its absence, in an isolated quantum system \cite{Rigol_review,Deutsch_review}. Whereas generic quantum many-body models satisfying the eigenstate thermalization hypothesis (ETH) are expected to thermalize at sufficiently long times \cite{Deutsch1991,Srednicki1994,Rigol_2008}, examples abound where interacting systems do not seem to thermalize for all accessible evolution times. For instance, many-body localization (MBL) has been argued to exist in the presence of a disorder potential \cite{Basko2006,Gornyi2005,Nandkishore_review,Abanin_review,Alet_review} or a sufficiently strong Stark potential in an otherwise translation-invariant system \cite{vanNieuwenburg2019,Schulz2019}. In the context of gauge theories, disorder-free localization can arise due to the plethora of conserved local constraints when the initial state spans an extensive number of gauge superselection sectors \cite{Smith2017,Brenes2018,Halimeh2022TDFL},
and the presence of such a substructure can be diagnosed using spectral signatures \cite{chakraborty2022disorder,chakraborty2022spectral}.
In certain nonintegrable models, initial states prepared in a cold subspace of quantum many-body scars---nonthermal eigenstates with anomalously low bipartite entanglement entropy and roughly equal energy spacing---lead to long-lived oscillations in the dynamics of local observables persisting beyond all relevant timescales and avoiding thermalization \cite{Bernien2017,Surace2020,Turner2018,Moudgalya2018}.

Recently, an exciting paradigm of ergodicity breaking known as Hilbert space fragmentation (HSF) has emerged, first reported for systems with  dipole conservation \cite{Sala2020,Khemani2020} where HSF can be fully characterized by nonlocal integrals of motion \cite{Rakovszky2020}. More generally, it was found that this phenomenon consists of the Hilbert space fragmenting into an exponentially large number of invariant subspaces resulting from an exponentially large number of commutant algebras \cite{Moudgalya2021}. HSF has received significant recent attention, with theoretical works demonstrating its presence in models with strict confinement \cite{DeTomasi2019,Yang2020fragmentation}, including gauge theories with a topological $\theta$-term \cite{Yang2020fragmentation,Desaules2023Ergodicity}, and it has been the focus of several recent ultracold-atom experiments \cite{Scherg2021,Kohlert2023exploring}.
A coupling between HSF and spatial fragmentation \cite{Patil2020Hilbert} can allow for
real space fragmentation and this can be exploited to control the spread of entanglement.

The sensitivity of HSF to perturbations has been highlighted since its introduction, where terms that violate local constraints have been shown to lead to eventual thermalization at sufficiently long times \cite{Sala2020,Moudgalya2022}. This sensitivity of HSF is disadvantageous from a fundamental point of view and for possible applications in quantum technology, and therefore it is important to investigate means of protecting dynamical features of HSF. In this paper, we explore the concept of quantum Zeno dynamics (QZD) \cite{facchi2002quantum,facchi2004unification,facchi2009quantum,burgarth2019generalized} in extending the lifetime of HSF-induced phenomena. In particular, we expand upon a recently proposed model \cite{Patil2020Hilbert}
that exploits HSF and spatial symmetry to engineer preservation of a perfectly disentangled structure for a large class of initial product conditions. 
This leads to no entanglement developing under Hamiltonian dynamics for particular spatial regions, which is a unique aspect of this model not found in general models exhibiting HSF. Although the model is fine-tuned, we show here that engineering large energy-scale separations by adding simple local interaction terms is able to thwart the growth of entanglement entropy (EE) due to perturbations that would otherwise render the model generically chaotic. Our results have implications in two important ways. First,
it illuminates the effect of a simple symmetry on the avoidance of thermalization. Second, from a quantum-information perspective, it outlines a method of protecting against unwanted entanglement buildup between different spatially separated parts of a quantum circuit, which can be useful in error mitigation.

The rest of the paper is organized as follows: We begin in Sec.~\ref{sec:model} by describing the model and the mechanism by which it prevents spread of entanglement entropy. This is followed
by a discussion of the importance of local conserved quantities and a spatial mirror symmetry, and we show numerically the effect of breaking the latter. In Sec.~\ref{sec:QZE} we discuss simple perturbations which break the local conservation and introduce a growth of entanglement entropy (EE), and propose a simple protection scheme based on QZD to suppress this growth. We justify our choice by studying the structure of the energy spectrum and show numerically that its strength controls parametrically the value and lifetime of a stable and suppressed EE plateau. This leads to scaling relations between the value of EE, the
perturbation strength, and the coefficient of the protection term, and we present analytic arguments from perturbation theory to justify them in Sec.~\ref{sec:PT}. We conclude and discuss future directions in which our results can be extended in Sec.~\ref{sec:conc}.

\section{Unperturbed Model $\hat{H}_0$}\label{sec:model}

\begin{figure}[t!]
\includegraphics[width=0.85\hsize]{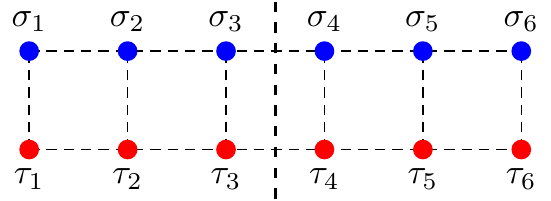}
\caption{Schematic picture of a spin system hosting two sets of identical spins ($\sigma$ and $\tau$) in a ladder geometry with a Hamiltonian given by Eq.~\eqref{eqH0}. Vertical dashed line
across the center represents the partition across which entanglement entropy is calculated.}
\label{fladder}
\end{figure}

We shall now discuss one of the simplest manifestations of the properties we require to prevent the spread of entanglement in a quantum many-body system, which is through a model of coupled Ising spins with Hamiltonian

\be\label{eqH0}
\hat{H}_0=-J\sum_{i} \big(\hat{\sigma}^z_i\hat{\sigma}^z_{i+1}+\hat{\tau}^z_i\hat{\tau}^z_{i+1}
+\hat{\sigma}^x_i\hat{\tau}^x_i\big),
\ee 
where the spin species are labelled by operators $\hat{\sigma}^a_i$ and $\hat{\tau}^a_i$ on site $i$ with $a=x,y,z$, and with eigenvalues $\sigma^a_i$ and $\tau^a_i$, respectively. The energy scale is set by $J=1$. This model was first discussed in Ref.~\cite{Patil2020Hilbert} for general lattices. For ease of numerical investigation, we shall here restrict our system to be a ladder with open boundary conditions, see Fig.~\ref{fladder}. Hamiltonian~\eqref{eqH0} conserves $\hat{\sigma}^z_i\hat{\tau}^z_i$ on each rung---$[\hat{H}_0,\hat{\sigma}^z_i\hat{\tau}^z_i]{=}0,\,\forall i$---leading to a fragmentation of the Hilbert space into $2^N$ fragments, where $N$ is the system size. Let us now restrict ourselves to one such sector, which can be represented as a particular assignment for $\sigma^z_i\tau^z_i$ such as $\ldots+++--+++++--+-----\ldots$, where $+(-)$ corresponds to ${\sigma^z_i\tau^z_i}=\pm 1$, and we shall denote an eigenvalue $\sigma^z_i$ or $\tau^z_i$ with a value $+(-)$ by $1(0)$ for ease of notation. For a rung, which hosts $+(-)$, $(\sigma^z_i,\tau^z_i)$ can thus be $00(01)$ or $11(10)$. 

Note that under an exchange of $\hat{\sigma}^a_i\leftrightarrow\hat{\tau}^a_i$,
the states in the $(\hat{\sigma}^z_i,\hat{\tau}^z_i)$ basis corresponding
to $\sigma^z_i\tau^z_i=+1$ are invariant, i.e., $00(11)\to00(11)$;
whereas those for $\sigma^z_i\tau^z_i=-1$ are interchanged, i.e.,
$01(10)\to10(01)$. Now note that the term $\hat{H}_d=-J\sum_j\big(\hat{\sigma}^z_i\hat{\sigma}^z_{i+1}+\hat{\tau}^z_i\hat{\tau}^z_{i+1}\big)$ of Hamiltonian~\eqref{eqH0} is diagonal
in the $(\hat{\sigma}^z_i,\hat{\tau}^z_i)$ basis, and is
symmetric under $\hat{\sigma}^a_i\leftrightarrow\hat{\tau}^a_i$.
This is the crucial property that we will exploit, which implies
that our argument should hold for any diagonal operator with this property. Consider now neighboring rungs 
$i$ and $i+1$ that
host $+-$ as their respective assignment of the conserved
quantity. This would imply that the matrix elements of the
inter-rung diagonal interaction,
$\braket{(00)_i(01)_{i+1}|\hat{H}_d|(00)_i(01)_{i+1}}$ and
$\braket{(00)_i(10)_{i+1}|\hat{H}_d|(00)_i(10)_{i+1}}$,
are equivalent under $\hat{\sigma}^a_i\leftrightarrow\hat{\tau}^a_i$,
which is a symmetry of the full Hamiltonian.
This means that there is effectively no interaction between
neighboring rungs that host different values for $\sigma^z\tau^z$.
As such, the
effective system within this sector breaks up into smaller
spatial segments given by the corresponding arrangement of
${\sigma^z_i\tau^z_i}$.
The Hamiltonian in this sector can now be written as
\begin{equation}
\hat{H}_\mathrm{sector}=-J\sum_s\bigg[\sum_{\braket{i,j}\in s}\big(\hat{\sigma}_i^z\hat{\sigma}_j^z
+\hat{\tau}_i^z\hat{\tau}_j^z\big)+\sum_{i\in s}\hat{\sigma}_i^x\hat{\tau}_i^x\bigg],
\end{equation}
where $s$ labels the different strings into which the sector is broken.
For example, a sector labelled by $+---++--$ would be made up of
four noninteracting strings with one site, three sites, two sites,
and two sites respectively.

Furthermore, the spatial decoupling discussed above leads to a $\mathbb{Z}_2$ Ising
symmetry associated with each string. This implies that the
sector breaks up further into $2^{N_s}$ subsectors, where $N_s$
is the number of strings. One can construct the conserved
quantities associated with these subsectors using the methodology
developed in Ref.~\cite{Moudgalya2021}, and these correspond to
products of $\hat{\sigma}^x_i\hat{\tau}^x_i$ on a string and projectors to the
domain wall which forces the $+-$ pattern at the ends of the
string of rungs.

\begin{figure}[t!]
\includegraphics[width=\hsize]{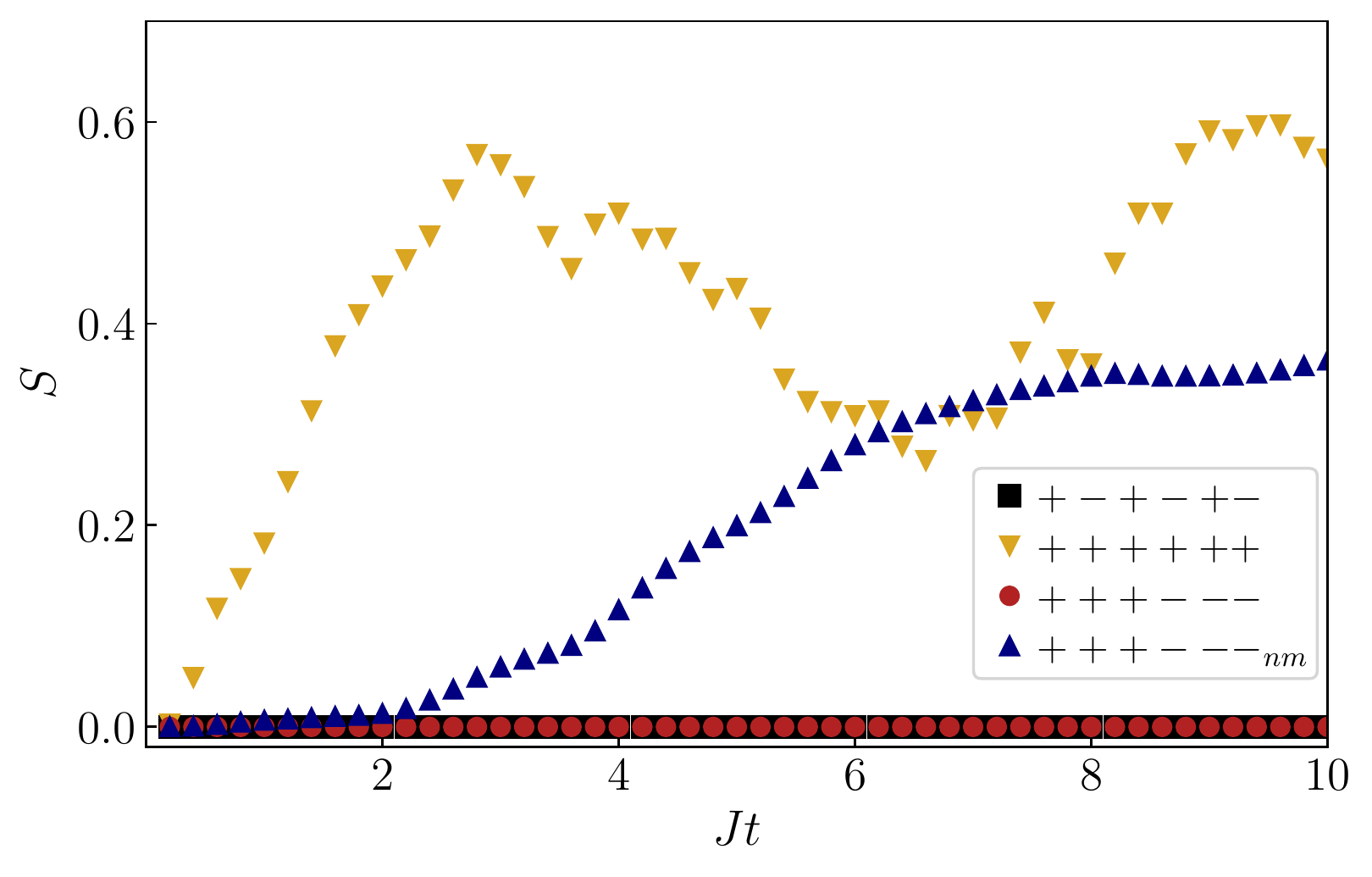}
\caption{Growth of the mid-chain EE for $\hat{H}_0$ for various sectors
labelled by the set of eigenvalues $(\pm 1)$ 
of $\hat{\sigma}^z_i\hat{\tau}^z_i$. 
Sectors with differing values across the middle develop no
entanglement under Hamiltonian time evolution (see text).
The subscript $nm$ signifies ``no mirror'', i.e., the system without a mirror symmetry,
which does not allow for a protection of information (see text).}
\label{figee0}
\end{figure}

We can now consider a product
initial state restricted to such a sector. Upon undergoing Hamiltonian dynamics, the corresponding growth of EE is now going to be limited to small spatial segments, and genuine volume-law EE can never develop. We verify this numerically for a six-rung ladder corresponding to $12$ spins, see Fig.~\ref{fladder},
by considering
random product states in sectors such as $+-+-+-$ and $+++---$, and find that
the half-chain (corresponding to a $+-$ wall)
EE to be exactly zero for all times, as shown in Fig.~\ref{figee0}.
We generate our random product
states, as done in Ref. \cite{kim2013ballistic} by choosing a random $(\theta_i,\phi_i)$ for rung number $i$,
and initializing the $(\sigma,\tau)$ pair on that rung as
$\cos\theta_i\ket{00}+\sin\theta_ie^{i\phi_i}\ket{11}$ if the
sector constraint is $\sigma^z_i\tau^z_i=+1$ on that rung or as $\cos\theta_i\ket{01}+\sin\theta_ie^{i\phi_i}\ket{10}$ if $\sigma^z_i\tau^z_i=-1$.

In contrast, a sector such as  
$++++++$ or $++++--$ develops partial half-chain
EE in the long-time limit. For the rest of this
manuscript, we shall study only the half-chain EE.
To emphasize the importance of the $\hat{\sigma}^a_i\leftrightarrow\hat{\tau}^a_i$ (mirror across rungs) symmetry, we show the dynamics of the EE under Hamiltonian~\eqref{eqH0} with the additional
$-0.1\sum_{\braket{i}} \hat{\tau}^z_i\hat{\tau}^z_{i+1}$, which breaks this mirror symmetry. Using the sector $+++---$
for our simulation, we
find, as shown in Fig.~\ref{figee0}, that there is a growth of EE that was not present
for the symmetric Hamiltonian. As such, the requirement for the $\hat{\sigma}^a_i\to\hat{\tau}^a_i$ mirror symmetry is
absolute, i.e., if we lose this property, every sector behaves as a generic many-body system, which builds volume-law EE in the long-time limit. However, this simply implies the requirement of a spatial symmetry, which is generally expected for periodic crystals.

\section{Protection with quantum Zeno dynamics}\label{sec:QZE}

\begin{figure}[t!]
\includegraphics[width=0.75\hsize]{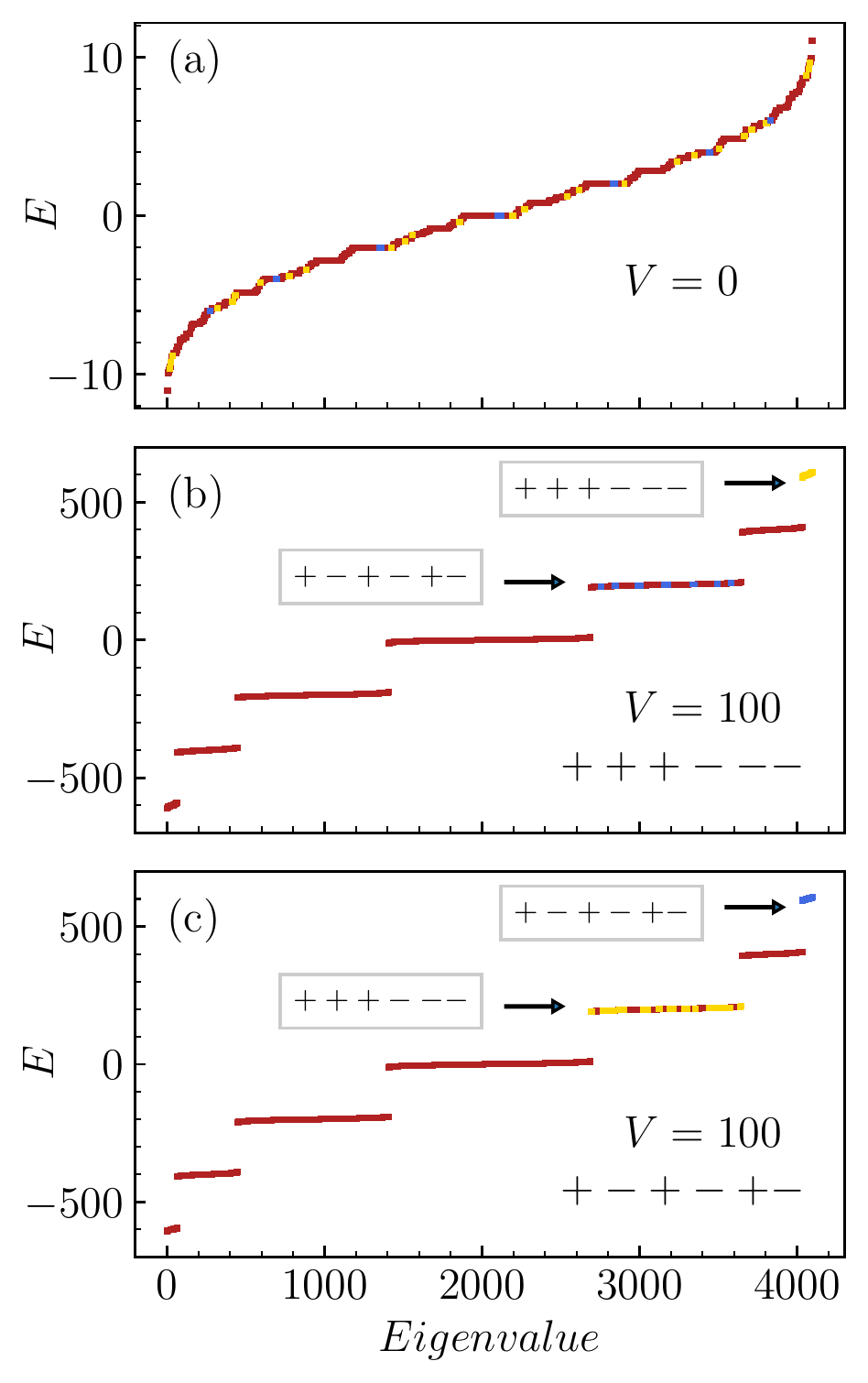}
\caption{Energy spectrum of $12$-site system ($L=6$ rungs) with 
$\lambda =0$,
(a) $V=0$, (b) $V=100$ with $c_i \in \{+1,+1,+1,-1,-1,-1\}$,
(c) $V=100$ with $c_i \in\{+1,-1,+1,-1,+1,-1\}$. The yellow and blue colors correspond to
states in the sectors $+++---$ and 
$+-+-+-$, 
respectively. For $V=0$, it can be clearly seen that there is
no energetic separation between various sectors, and the dynamics
is restricted only by the conserved quantities, whereas for
$V\gg J=1$, the large energy separation ensures protection against
perturbations with the correct sequence $c_i$. Note that the spectra shown in panels (b) and (c) show little difference to the case with a finite nonzero $\lambda\ll V$}
\label{figspec}
\end{figure}

Let us now consider perturbations $\hat{H}_1$ that are off-diagonal in the $(\hat{\sigma}^z,\hat{\tau}^z)$ basis. These terms do not affect the perfect mirror symmetry of $\hat{\sigma}^z_i\leftrightarrow\hat{\tau}^z_i$, but destroy the conservation of $\hat{\sigma}^z_i\hat{\tau}^z_i$ on each rung, and thereby enhance the quantum dynamics and destroy HSF. This occurs due to a loss of sector structure, and thus a leakage of EE due to a sector with the $\sigma^z_i\tau^z_i$ assignment of the
form $\ldots+-\ldots$ being able to transition to one with an assignment of $\ldots++\ldots$, where entanglement can build up between the two relevant sites. In the following, we will consider two such perturbations, both of which mix all sectors in such a way, leading to the EE of a random state \cite{page1993average} in the Hilbert space in the long-time limit.

We can now employ the concept of quantum Zeno dynamics (QZD) \cite{facchi2002quantum,facchi2004unification,facchi2009quantum,burgarth2019generalized} in order to protect the constrained dynamics arising originally due to HSF and mirror symmetry in $\hat{H}_0$. QZD allows us to project the Hamiltonian dynamics into the sector of choice, thus negating the effect of the perturbation $\hat{H}_1$. This can be achieved by adding a \textit{protection} term $\hat{H}_V$ diagonal in the $(\sigma^z,\tau^z)$ basis with strength $V$, which is the largest energy scale in our Hamiltonian. This term is chosen to have a spectrum that forms plateaus, such that all the $(\sigma^z,\tau^z)$ basis vectors in a sector generate the same expectation value for $\hat{H}_V$ (spectrum shown in Fig.~\ref{figspec}).
Note that this implies that each sector corresponds to only
one plateau, but each plateau can contain many sectors. Now the
energy separation between different plateaus leads to a
suppression of inter plateau dynamics \cite{facchi2009quantum},
and thus a suppression of inter-sector dynamics, provided
the sectors are in different energy plateaus.
Note that the suppression of inter-sector dynamics is perfect only for
$V\to\infty$, and that for any finite value of $V$, we should expect
some small renormalized value for the EE, and that it would last only
up to a finite time, till which point the dynamics accumulates enough
leakage out of the sector, and leads to thermalization. Indeed, from QZD one obtains an effective Zeno Hamiltonian $\hat{H}_\text{Z}=\hat{H}_0+\sum_\mathbf{S}\hat{\mathcal{P}}_\mathbf{S}\hat{H}_1\hat{\mathcal{P}}_\mathbf{S}$ that faithfully reproduces the HSF dynamics up to timescales at the earliest linear in $V$, where $\hat{\mathcal{P}}_\mathbf{S}$ is the projector onto the quantum Zeno subspace $\mathbf{S}$ \cite{facchi2002quantum}. The Zeno Hamiltonian further reduces to $\hat{H}_\text{Z}=\hat{H}_0+\hat{\mathcal{P}}_{\mathbf{S}_0}\hat{H}_1\hat{\mathcal{P}}_{\mathbf{S}_0}$, when the initial state is prepared in a given Zeno subspace $\mathbf{S}_0$.

\begin{figure}[t]
\includegraphics[width=\hsize]{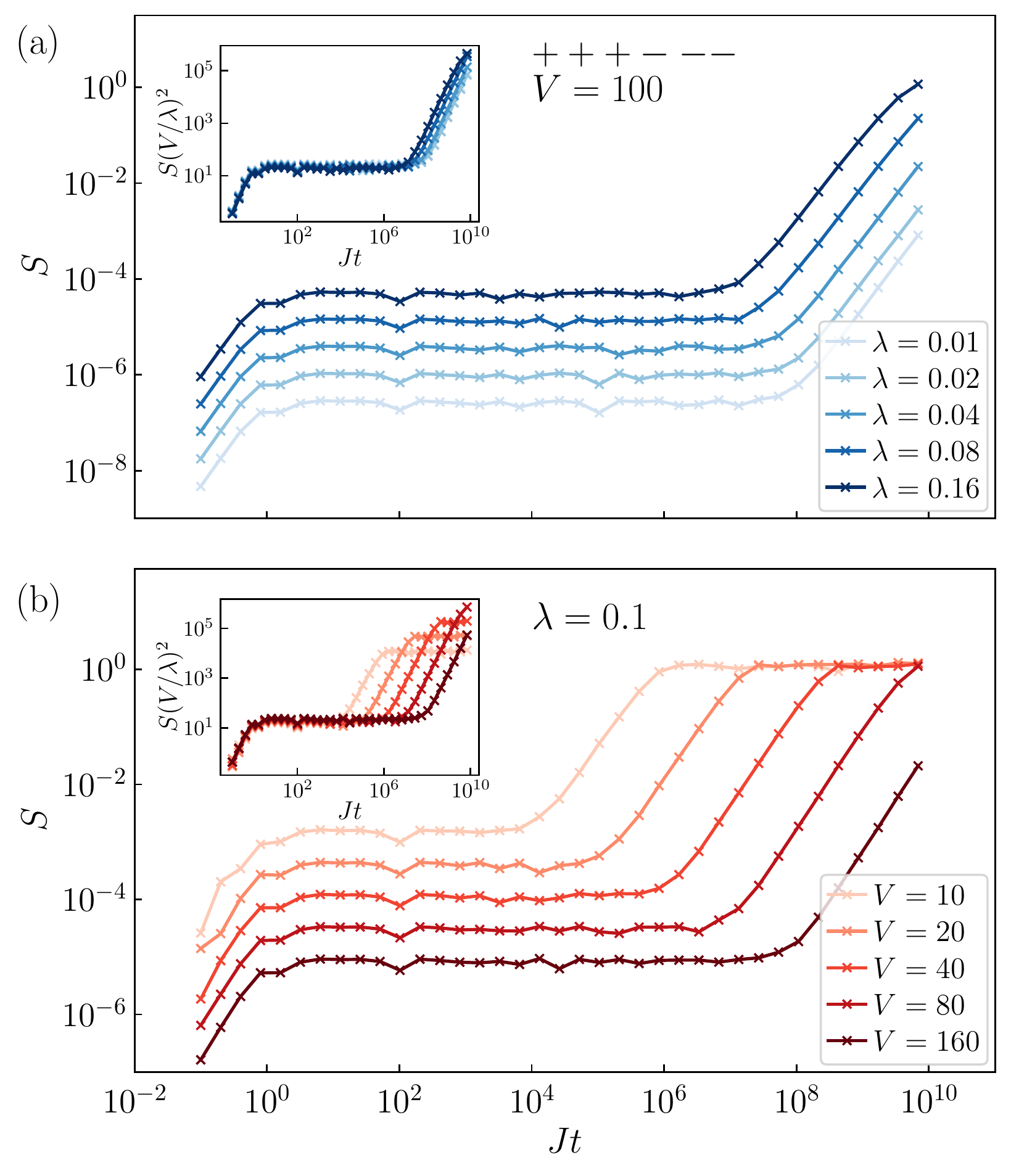}
\caption{Entanglement entropy growth for a product initial state
in the $+++---$ sector with
$c_i\in\{+1,+1,+1,-1,-1,-1\}$ in the presence of the transverse-field perturbation.
(a) Varying $\lambda$ for $V=100$ or
(b) varying $V$ for $\lambda =0.1$, shows a
prethermal plateau in the EE whose value and duration is controlled
by $(V,\lambda)$ leading to a controllable suppression of the EE.
The insets show the EE rescaled by $(V/\lambda)^2$, signifying the value of the EE prethermal plateau $\propto\lambda^2/V^2$.}
\label{fig3a1}
\end{figure}

To design a protection term particularly well-suited to
prevent the growth of entanglement starting from a specific
sector, we must bias it towards the configuration of $\sigma^z_i\tau^z_i$ corresponding to this sector.
This can be done by choosing $\hat{H}_V= V \sum_i c_i\hat{\sigma}^z_i\hat{\tau}^z_i$,
where $c_i=\pm 1$, depending on the sector assignment.
For example, for an assignment of $+++---$, we would have
$c_i\in\{+1,+1,+1,-1,-1,-1\}$. This choice of $c_i$ completely energetically
isolates the sector---rendering it a unique Zeno subspace $\mathbf{S}_0$---as all other assignments must necessarily frustrate one of the $c_i$, leading to a shift to a different plateau.
The next closest sector, created by toggling a single $+\leftrightarrow-$, is thus separated in the spectrum by an energy gap of $2V$. Figure~\ref{figspec}
shows the energy spectrum for $\hat{H}_0$ and
$\hat{H}_V$, with two different target sectors highlighted. First consider Fig.~\ref{figspec}a,
where the spectrum of $\hat{H}_0$
is shown. Here we see that there is no energy separation between
different sectors and thus sectors are protected against mixing only by
the local conservation of $\sigma^z_i\tau^z_i$. 
The introduction of a large $V$ leads to
a clear separation of the spectrum into plateaus, with the targeted
sector floating to the top of spectrum and being well separated from
all other sectors. However, nontargeted sectors remain in the middle
of the spectrum, sharing energy levels with other sectors. These
features can be seen in Fig.~\ref{figspec}b,c. In this Section,
we find that such an energy separation translates dynamically into a prethermal regime
due to projective quantum Zeno dynamics, which can be seen as a 
plateau in EE with a tunable value and lifetime.

\begin{figure}[t]
\centering
\includegraphics[width=\hsize]{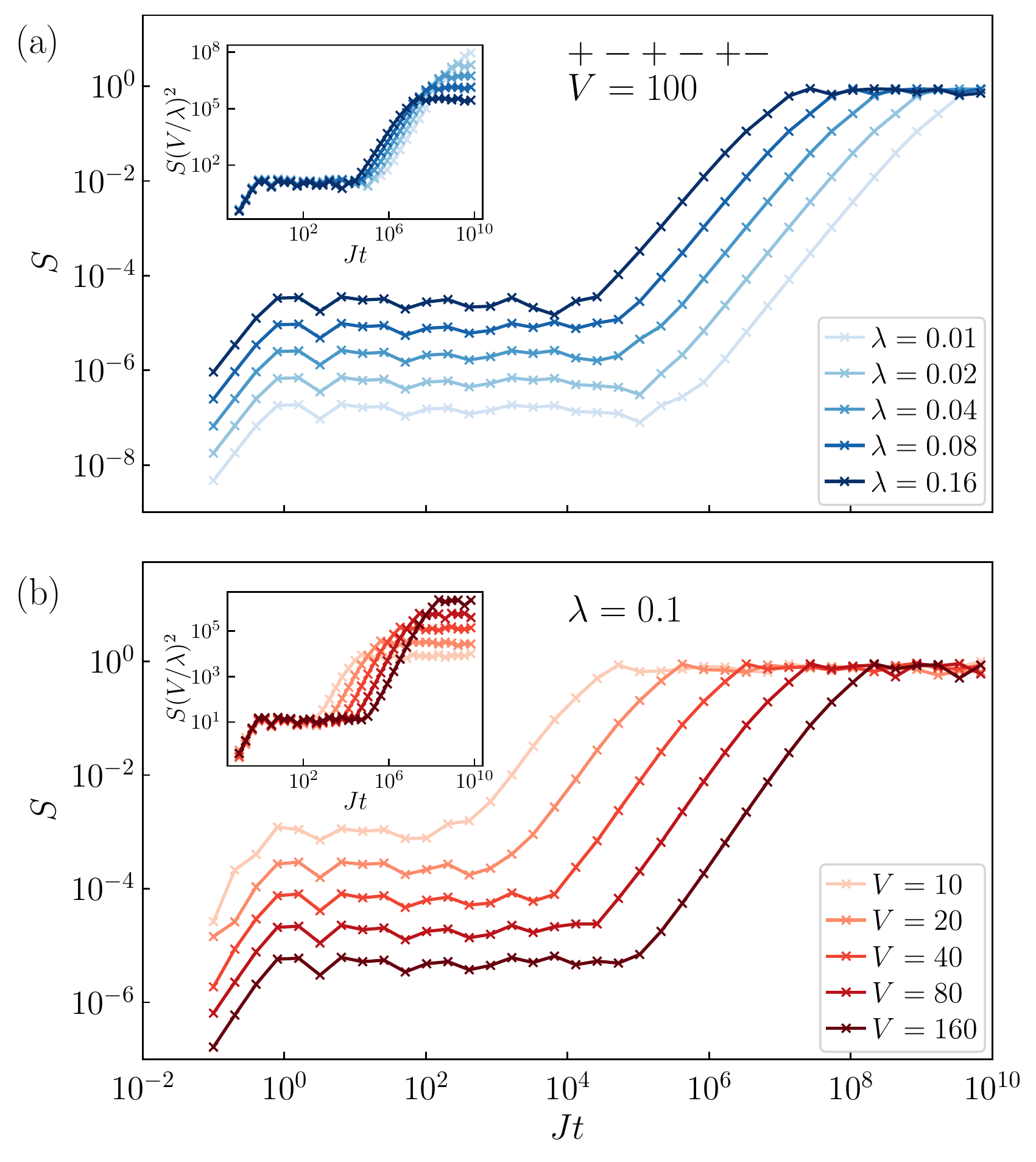}
\caption{Entanglement entropy growth for a product initial state
in the sector $+-+-+-$ with
$c_i\in\{+1,-1,+1,-1,+1,-1\}$ in the presence of the transverse-field perturbation.
(a) Varying $\lambda$ for $V=100$ or (b) varying $V$ for
$\lambda =0.1$ show similar features as Fig.~\ref{fig3a1}. The insets show the rescaled EE, indicating that the EE settles into a prethermal plateau $\propto\lambda^2/V^2$.}
\label{fig3a2}
\end{figure}

In Secs.~\ref{S3a} and~\ref{S3b},
we consider two different forms for the perturbation $\hat{H}_1$, and
numerically investigate their effect on EE growth in different
sectors under specific protection for these sectors.
The numerical results in these subsections show the occurrence 
of the EE plateau discussed above.
In Sec.~\ref{S3c}, we examine the lifetime of this prethermal
plateau, and show that its qualitative dependence on $V$ depends
on the structure of the given sector that we choose to protect.
In Sec.~\ref{S3d},
we investigate the degree of protection that
a particular choice of terms can offer to sectors it is
not explicitly designed to protect, and its dependence on the type
of perturbation used. We close this section by discussing our
results in the thermodynamic limit in Sec.~\ref{S3e}, where
we use the time-dependent variational principle (TDVP) algorithm \cite{haegeman2016unifying, haegeman2011time,vanderstraeten2019tangent, leviatan2017quantum} to simulate larger system sizes.

\subsection{Transverse-field perturbation}\label{S3a}

\begin{figure}[t]
\centering
\includegraphics[width=\hsize]{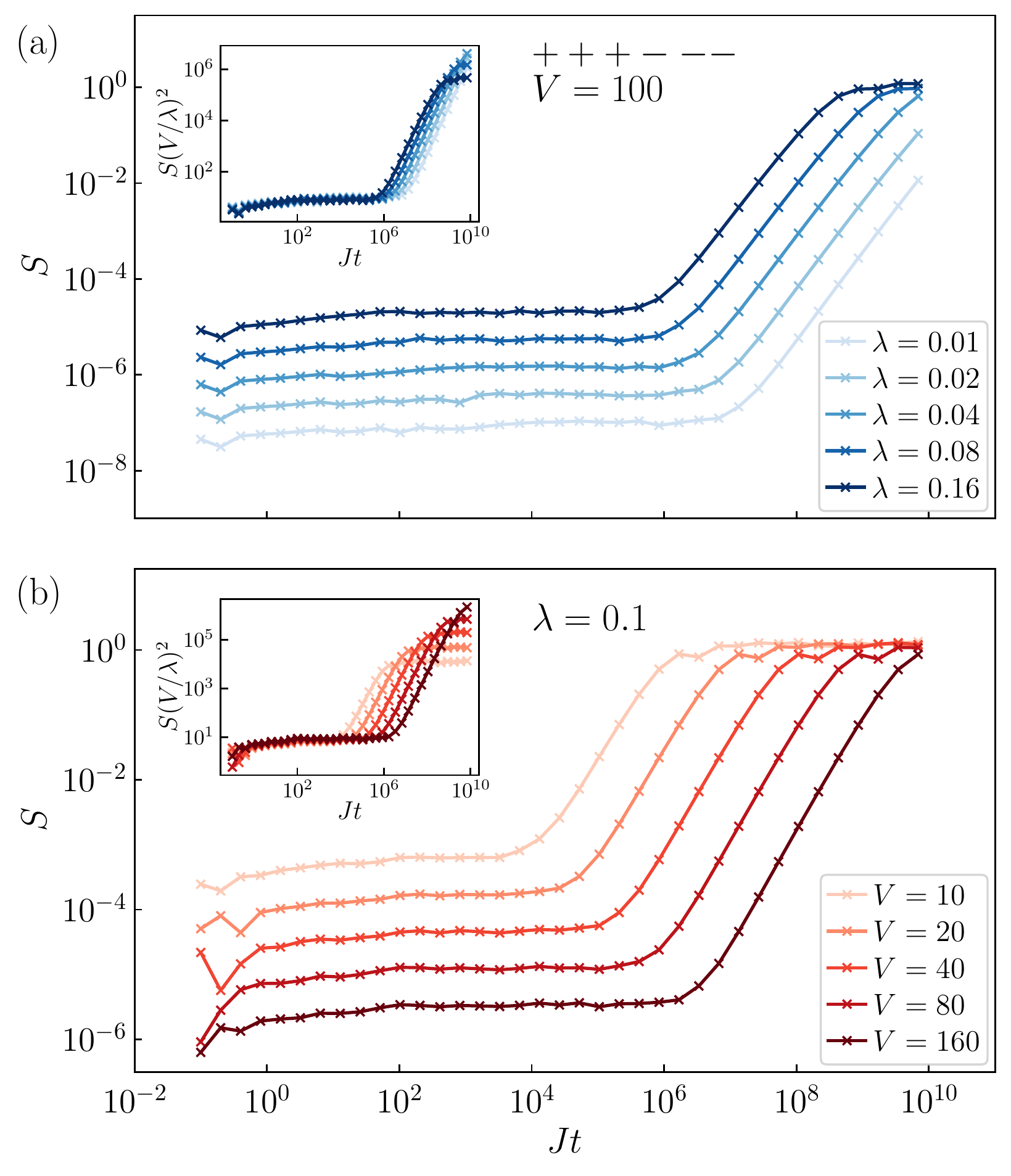}
\caption{Entanglement entropy growth for a product initial state in the
sector $+++---$ with the Heisenberg perturbation and with $c_i\in\{+1,+1,+1,-1,-1,-1\}$. (a)
Varying $\lambda$ for $V=100$ or
(b) varying $V$ for $\lambda =0.1$, shows a controlled prethermal EE plateau $\propto\lambda^2/V^2$. The insets show the rescaled EE.}
\label{fig3b1}
\end{figure}

Let us examine the quench dynamics under $\hat{H}=\hat{H}_0+\hat{H}_1+\hat{H}_V$ with a simple global transverse-field perturbation
$\hat{H}_1=\lambda\sum_i(\hat{\sigma}^x_i+\hat{\tau}^x_i)$, which violates the
local conservation of $\sigma^z_i\tau^z_i$ by toggling its value on a single rung from $+\leftrightarrow -$. This allows transitions, for e.g., from ``insulating'' sectors of the type $+++---$ to ``conducting''
ones of type $++++--$. We study the suppression of EE for initial
product states belonging to the
sectors $+++---$ and $+-+-+-$, both of which prevent entanglement
buildup across the center of the chain in the ideal case of $\lambda=0$, by implementing protection terms
specific to these sectors as discussed above.
The results for the half-chain EE from exact diagonalization of a $12$-spin ($6$-rung) system are shown in Figs.~\ref{fig3a1}
and \ref{fig3a2} for a range of $\lambda$ and $V$ values. Note that
we see a plateau with a small value of EE from intermediate to
large values of $Jt$ that scales $\propto\lambda^2/V^2$. This plateau eventually leads to a slow growth
of the EE to the saturation value, which is expected to be proportional
to the system volume \cite{page1993average}. The value of the EE at the plateau can be clearly
seen to decrease quadratically with increasing values of $V/\lambda$. This behavior
can be intuitively understood by considering the $V$ and $\lambda$ terms
to have opposing effects, i.e., the perturbation promotes a spread of the wave function into other sectors, in which entanglement can be built
up, whereas the protection term projects it back into the sector of interest according to QZD.
The insets in Figs.~\ref{fig3a1} and \ref{fig3a2}
show a data collapse by rescaling the EE
by $(V/\lambda)^2$. An analytic understanding of this scale is provided by a perturbation theory analysis in Sec.~\ref{sec:PT}. Note that there
is no significant difference between the behavior of the EE for the
sectors $+++---$ (Fig.~\ref{fig3a1}) and the sector $+-+-+-$ (Fig.~\ref{fig3a2}),
as in each case the protection term is appropriately chosen.

\subsection{Heisenberg perturbation}\label{S3b}

\begin{figure}[t]
\centering
\includegraphics[width=\hsize]{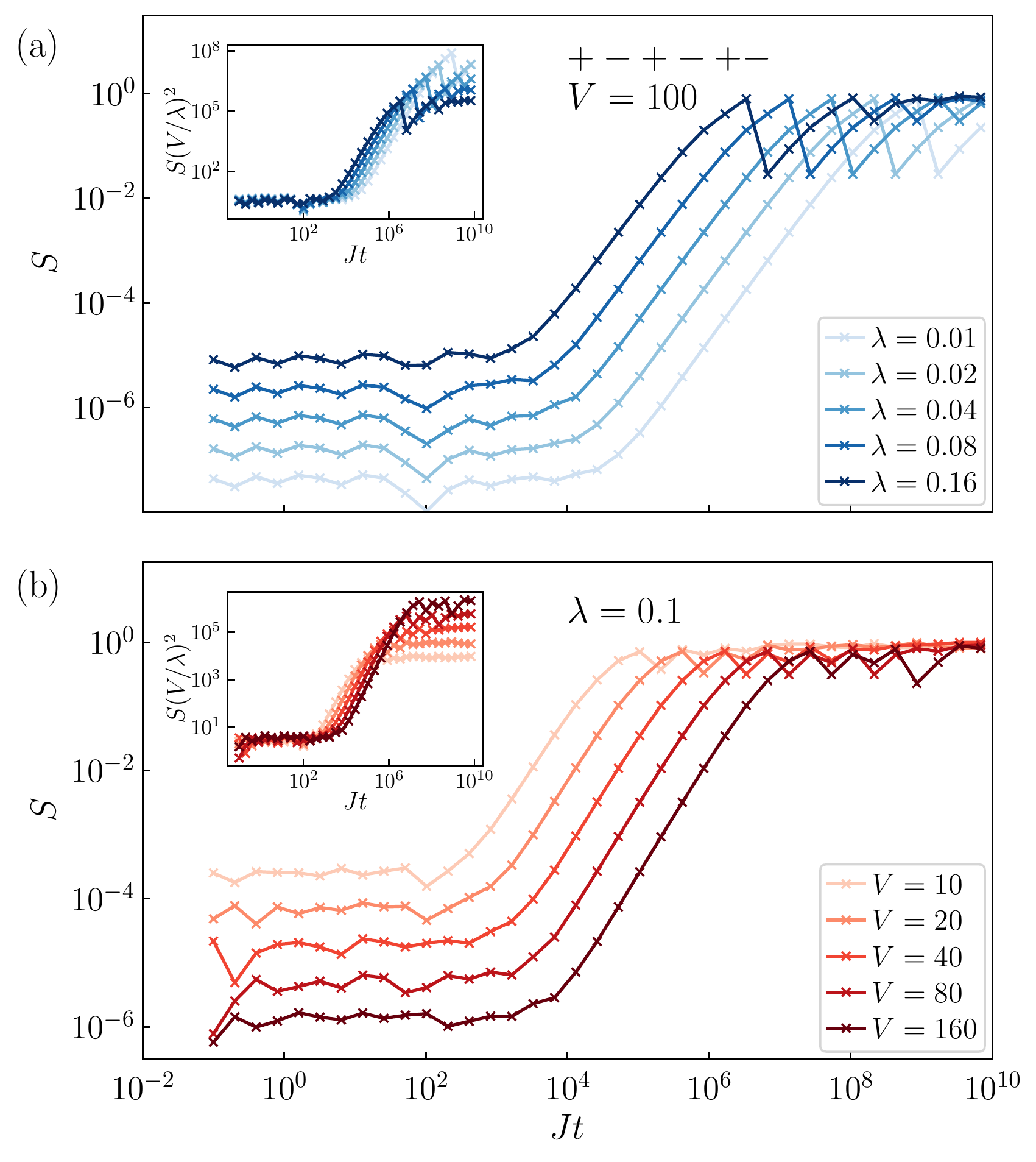}
\caption{Entanglement entropy growth for a product initial state in the
sector $+-+-+-$ and $c_i\in\{+1,-1,+1,-1,+1,-1\}$ in the presence of a Heisenberg perturbation. (a) Varying $\lambda$ for $V=100$ or
(b) varying $V$ for $\lambda =0.1$ shows a controlled prethermal EE plateau $\propto\lambda^2/V^2$. The insets show the rescaled EE.}
\label{fig3b2}
\end{figure}

For a perturbation of the form
$\hat{H}_1=\lambda\sum_i(\hat{\sigma}^x_i\hat{\sigma}^x_{i+1}+\hat{\sigma}^y_i\hat{\sigma}^y_{i+1}+\hat{\tau}^x_i\hat{\tau}^x_{i+1}+\hat{\tau}^y_i\hat{\tau}^y_{i+1})$, we find that the EE behaves in a manner similar to
the case of the transverse-field perturbation, as demonstrated in Figs.~\ref{fig3b1} and~\ref{fig3b2} for initial product states in the sectors $+++---$ and $+-+-+-$. This Heisenberg perturbation term acts by toggling the value of 
$\sigma^z_i\tau^z_i$ on neighboring rungs. We show in
Figs.~\ref{fig3b1} and~\ref{fig3b2}
the growth of the EE for various values
of $V$ and $\lambda$, where once again we find a convincing data collapse by rescaling the EE by $(V/\lambda)^2$. We therefore see that the protection term $\hat{H}_V$, for an appropriate sequence $c_i$, will stabilize HSF-induced constrained dynamics with the EE exhibiting a plateau $\propto\lambda^2/V^2$ up to times linear in $V$ in a worst-case scenario. This shows that the protection term is able to reliably stabilize the system against different types of perturbations.

\subsection{Lifetime of prethermal plateau}\label{S3c}
\begin{figure}[t]
\centering
\includegraphics[scale=0.5]{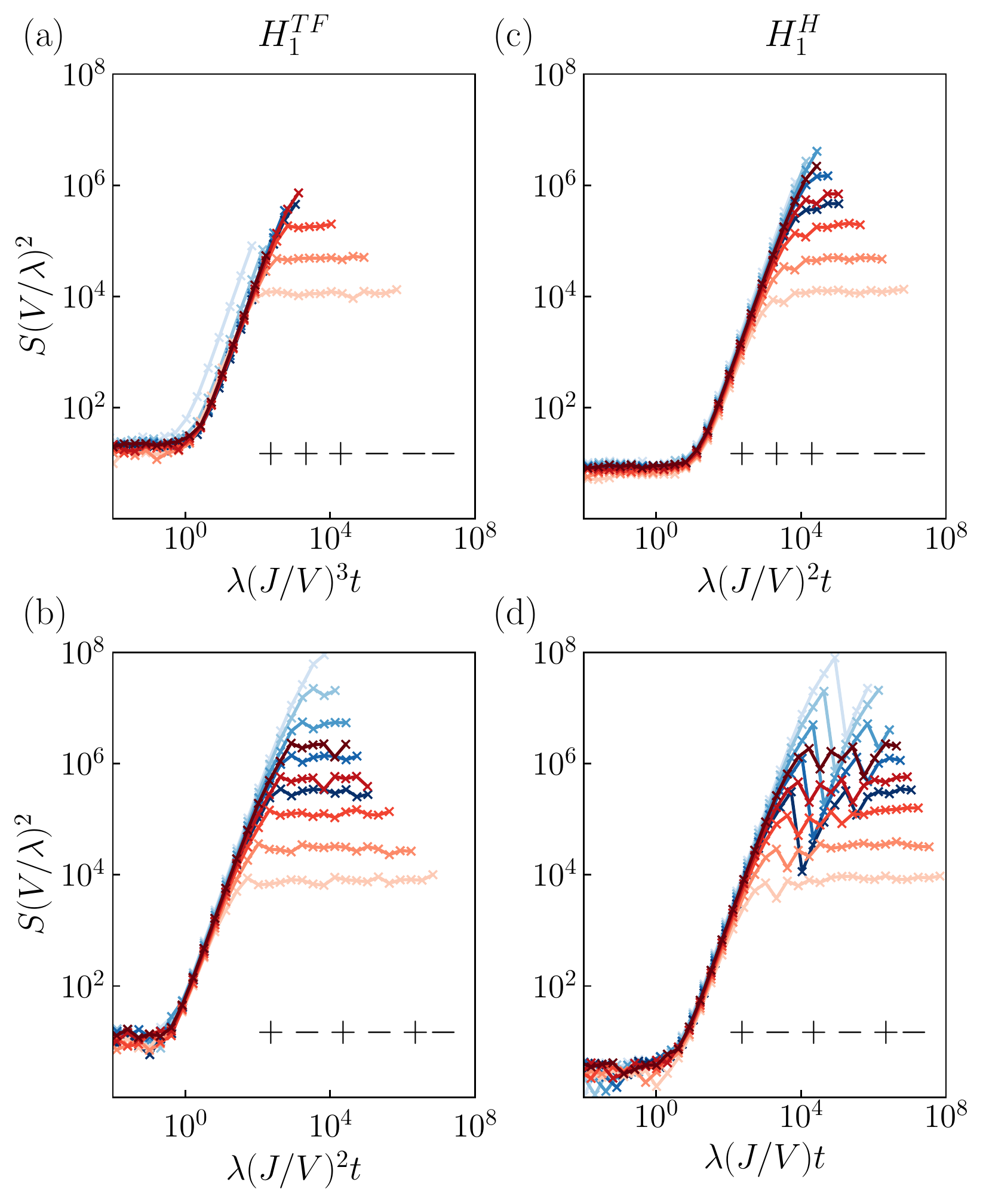}
\caption{Rescaled entanglement entropy growth with appropriate 
scaling in the presence of (a,b) transverse-field perturbation and (c,d) the 
Heisenberg perturbation when starting in (a,c) the $+++---$ sector or (b,d) the $+-+-+-$ sector. The blue and red curves are variations 
of $\lambda$ with fixed $V=100$  and variations of $V$ with fixed 
$\lambda=0.1$, respectively, with the same values used in Figs.~\ref{fig3a1}--\ref{fig3b2}.}
\label{fig3lt}
\end{figure}

Let us now take a closer look at the timescales of the EE plateaus shown in Figs.~\ref{fig3a1}--\ref{fig3b2}. As we have previously mentioned, a properly chosen sequence $c_i$ allows the protection term $\hat{H}_V$ to induce QZD dynamics that reliably and controllably stabilizes the EE into a plateau of value $\propto\lambda^2/V^2$ up to a timescale that is, in the worst-case scenario, linear in $V$ \cite{facchi2002quantum}. By properly rescaling the time axes, Fig.~\ref{fig3lt} illustrates the timescales of the EE plateaus resulting from the dynamics shown in Figs.~\ref{fig3a1}--\ref{fig3b2}. In the case of the transverse-field perturbation, we observe that when
we target the sector $+++---$, a rescaling of the time axis to $\lambda(J/V)^3t$ results in a convincing data collapse for
the complete range of $(\lambda,V)$ which we have studied, indicating a timescale of the EE plateau $\propto V^3/(J^3\lambda)$, which is significantly better than the linear-in-$V$ worst-case prediction from QZD. For the sector $+-+-+-$, we find that we must modify our scaling to $\lambda(J/V)^2t$, leading to a plateau timescale $\propto V^2/(J^2\lambda)$, also exceeding the QZD prediction. For the Heisenberg perturbation, the plateau timescale for the initial state in the sector $+++---$ is $\propto V^2/(J^2\lambda)$ and for the initial state in the sector $+-+-+-$ is $V/(J\lambda)$, implying a more adverse effect of this perturbation than its transverse-field counterpart. We note that such a difference in the Zeno timescales based on the type of error has also been reported in the context of lattice gauge theories \cite{halimeh2022enhancing}.

\subsection{Protection of non-targeted sectors}\label{S3d}
\begin{figure}[t]
\centering
\includegraphics[width=\hsize]{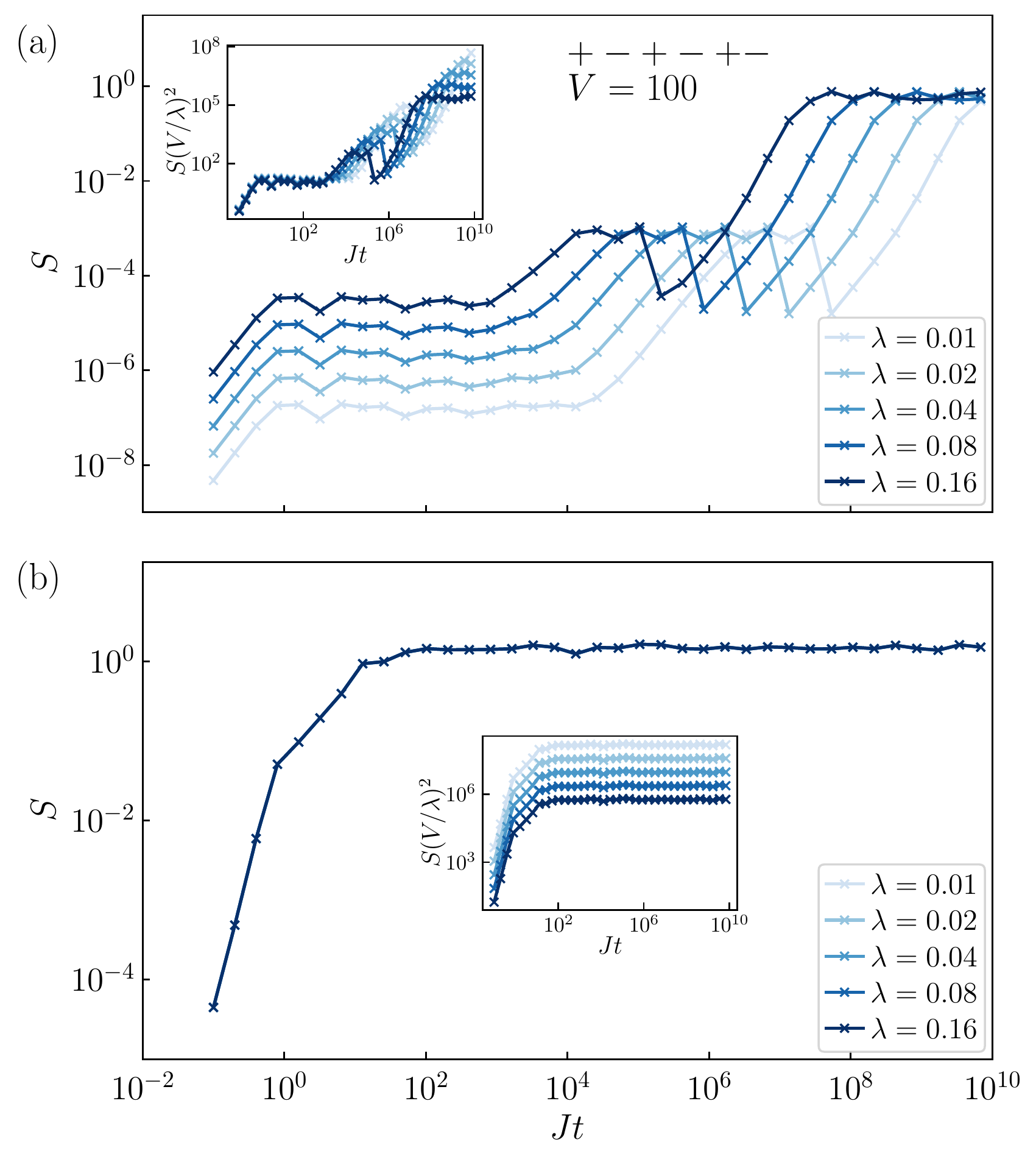}
\caption{Entanglement entropy growth for a product initial state in the
sector $+-+-+-$ with $c_i\in\{+1,+1,+1,-1,-1,-1\}$ for (a) the transverse-field perturbation, and (b) the Heisenberg perturbation with varying $\lambda$ for 
$V=100$. The insets show the rescaled EE. This shows that even the ``wrong'' sequence $c_i$ can still adequately suppress EE growth for certain types of perturbations, but not generically.}
\label{fig3d}
\end{figure}

Here we would like to examine the extent to which our protection term
can protect sectors for which it is not specifically designed to protect.
As seen in the spectrum shown in Fig.~\ref{figspec}c,
an example of such a sector
would be $+-+-+-$. We reiterate that the protection term 
corresponding
to the spectrum here contains the sequence $c_i\in\{+1,+1,+1,-1,-1,-1\}$, and thus does not follow the configuration of the sector we wish to protect.
However, it is still possible to prevent a growth of EE
for perturbations that are unable to connect sectors lying in the same energy plateau.
An example of such a perturbation is the transverse-field
term $\hat{H}_1=\lambda\sum_i(\hat{\sigma}^x_i+\hat{\tau}^x_i)$
discussed in Sec.~\ref{S3a}. The action of this perturbation
is to toggle $+\leftrightarrow -$ on a rung $i$. Under this
operation, $\braket{\hat{H}_V}$ changes by $\approx2Vc_i$, thus 
accessing a sector
which is well separated in energy. This leads to a suppression
due to Zeno dynamics, and a projection into the initial sector.
We verify this numerically in Fig.~\ref{fig3d}a, where we find
a plateau at short to intermediate times for the EE which scales
with $(\lambda/V)^2$. This is consistent with our results in
Secs.~\ref{S3a} and~\ref{S3b}.

In comparison,
under the application of the Heisenberg perturbation,
the initial sector $+-+-+-$ can transition to, e.g., $++--+-$.
This is in the same energy plateau, as our choice of
$c_i\in\{+1,+1,+1,-1,-1,-1\}$ leads to a cancellation in the
change in $\braket{\hat{H}_V}$ due to the double toggle. This can be seen by evaluating
$\sum_i c_i\sigma^z_i\tau^z_i$ for both sectors, and implies that the sector is 
unprotected. Furthermore, as the sector $++--+-$
has the same value of $\sigma^z_i\tau^z_i$ on either side of its center, the growth of EE is unrestricted, leading to
vanilla saturation to an EE proportional to system size,
with no dependence on $\lambda,V$. This is clearly seen in
Fig.~\ref{fig3d}b for a range of values of $\lambda$.

\subsection{Thermodynamic limit (TDVP results)}\label{S3e}

For the ladder system studied here, the only perturbations that can lead to a buildup of entanglement across the middle of the chain are of the form
$\pm\mp\leftrightarrow ++$ and $\pm\mp\leftrightarrow --$. 
Above, we show that the protection scheme controls the growth of entanglement entropy for small system sizes, with a clear scaling of the
prethermal plateau and its lifetime. However, it is important to
understand the effect of larger system sizes, as 
a larger Hilbert space may lead to a higher entropic pressure
on the barrier to EE growth.

\begin{figure}[t]
\centering
\includegraphics[width=\hsize]{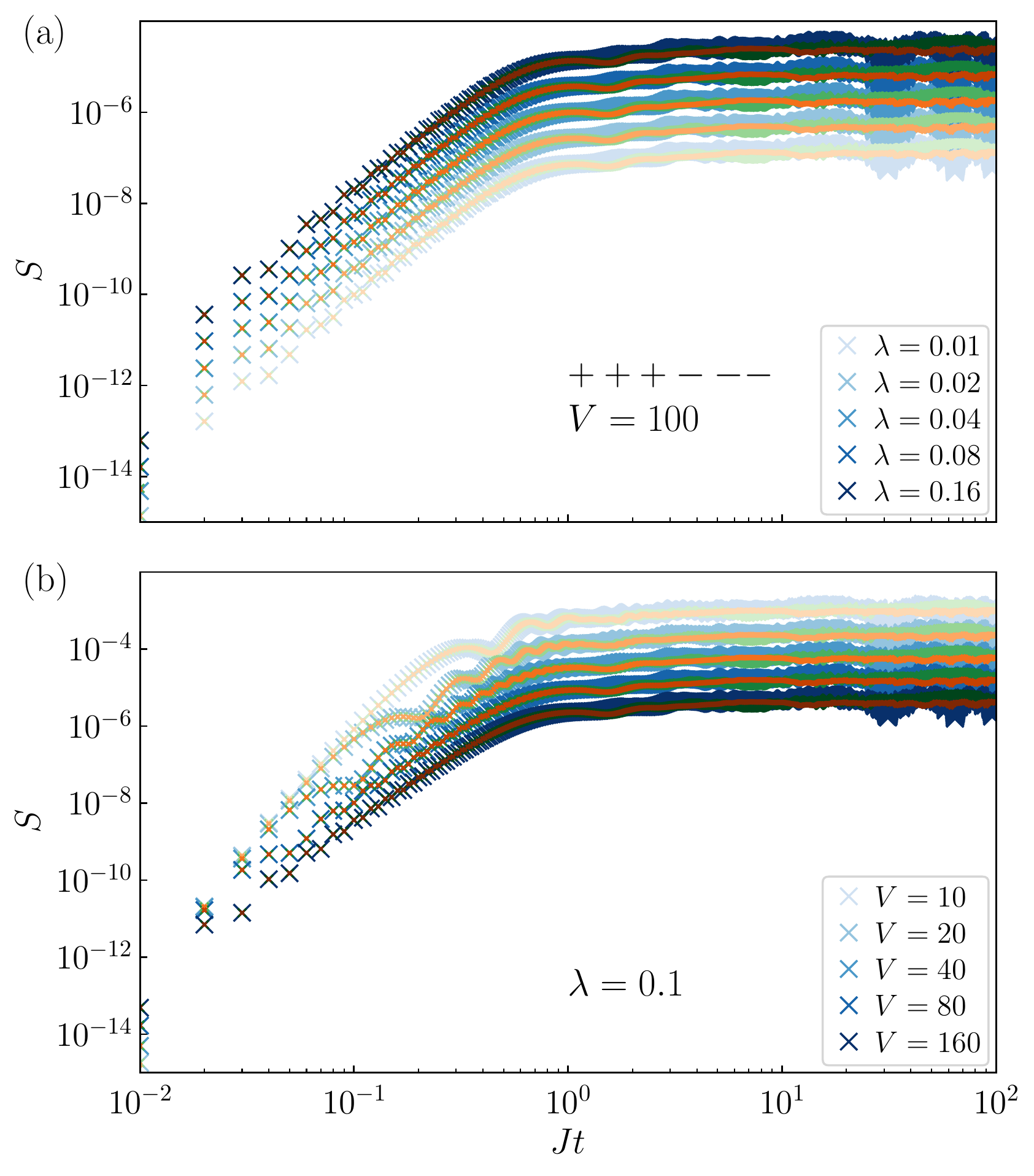}
\caption{Entanglement entropy growth for a product initial state in the
sector $\ldots+++---\ldots$ in the presence of the transverse-field perturbation with 
$c_i\in\{+1,+1,+1,-1,-1,-1\}$ for various system sizes calculated using TDVP.
(a) Varying $\lambda$ for $V=100$ or (b) varying $V$ for $\lambda
=0.1$, again shows suppression of the EE to a prethermal plateau $\propto\lambda^2/V^2$. Blue, green, and orange curves correspond to system sizes $N=12$, $20$, and $40$, respectively. The color shade gets darker with increasing values of $\lambda$ and $V$.}
\label{tdvp++}
\end{figure}

\begin{figure}[t]
\centering
\includegraphics[width=\hsize]{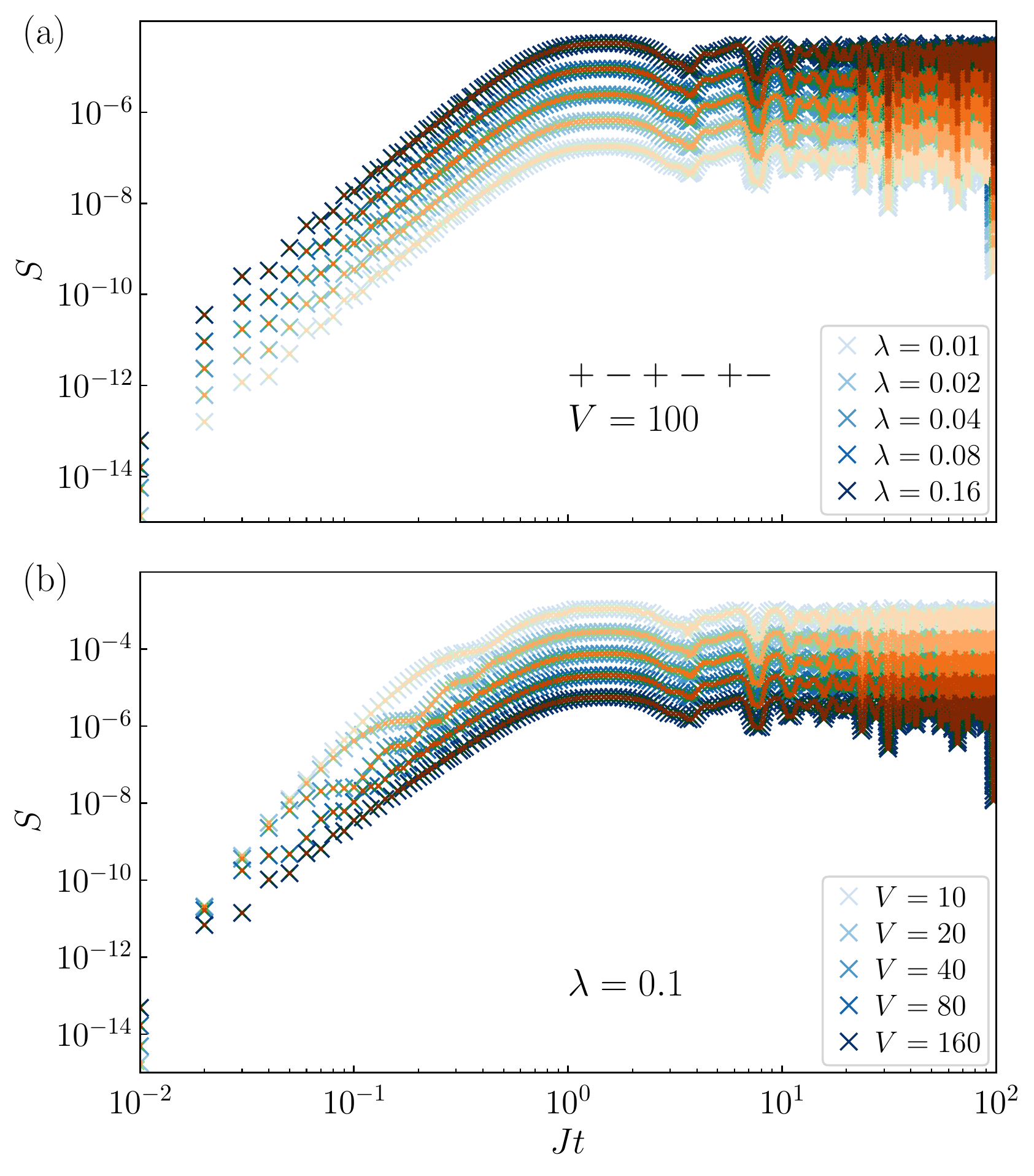}
\caption{Entanglement entropy growth for a product initial state in the
sector $\ldots+-+-+-\ldots$ in the presence of the transverse-field perturbation with 
$c_i\in\{+1,-1,+1,-1,+1,-1\}$ for various system sizes.
(a) Varying $\lambda$ for $V=100$ or (b) varying $V$ for
$\lambda=0.1$ again shows a suppression of the EE into a prethermal plateau $\propto\lambda^2/V^2$. Blue, green, and orange curves correspond to system sizes $N=12$, $20$, and $40$, respectively. The color shade gets darker with increasing values of $\lambda$ and $V$.}
\label{tdvp+-}
\end{figure}

To confirm that our results for the $12$-site system carry
over to the thermodynamic limit, we study the dependence on system size. To this end, we employ the time-dependent variational principle (TDVP) for matrix product states (MPS) \cite{haegeman2016unifying, haegeman2011time,vanderstraeten2019tangent, leviatan2017quantum}. In general, it is intractable to study the long-time evolution of an interacting quantum many-body system, as it requires a prohibitively large bond dimension that usually grows exponentially in evolution time \cite{Uli_review}. However, in our case, the protection term leads to constrained dynamics that requires a much smaller bond dimension, thus allowing us to access large system sizes. We perform the calculations for $N=12$, $20$, and $40$ for two different initial states perturbed by the transverse-field term and with an appropriate protection term. Figure~\ref{tdvp++} shows the result with one domain wall in the middle of the ladder and Fig.~\ref{tdvp+-} shows the results with an alternatingly repeating domain wall, similar to the initial states discussed above. The maximum bond dimension is kept to $D=200$ and maximum fidelity threshold $10^{-14}$ for $N=40$. The calculations are performed using ITensor software library \cite{itensor}. The results confirm that the entanglement growth in the middle of chain remains almost unaffected with increasing system size.

\section{Scaling relations from perturbative analysis}\label{sec:PT}
The numerical results indicate the existence of two universal scaling relations.
The first is
the collapse with $(\lambda/V)^2$ of the EE plateau developed at
intermediate to long
times, and the second is the time at which the EE begins to increase
again, which is seen to scale with the Hamiltonian parameters as
$t\propto V/(J\lambda)$ in a worst-case scenario. Here we only attempt to understand
the first scaling behavior, as it is tractable within first-order
perturbation theory, while the second scaling behavior is understood from QZD \cite{burgarth2019generalized}. The argument that we build in this
section follows broadly the following structure: At first-order perturbation, there are no contributions from within
the initial sector, thus all contributing states have a suppression
of $\mathcal{O}(\lambda/V)$. This translates to a contribution to the EE
of $\mathcal{O}(\lambda^2/V^2)$ for the plateau. We argue this in detail below.

We carry out our analysis by considering the smallest possible system
where we can study this behavior, namely a two-rung/four-site system.
To understand the scaling of the EE plateau, let us consider an
initial condition that belongs to the sector $+-$. The Hamiltonian in
this sector can be rewritten in terms of broken segments,
as defined in Sec.~\ref{sec:model}, as $\hat{H}_0=-\hat{\sigma}^x_1\hat{\tau}^x_1-
\hat{\sigma}^x_2\hat{\tau}^x_2$. The
eigenstates of this Hamiltonian are simply given by the product
states of eigenstates of $\sigma^x_{1(2)}\tau^x_{1(2)}$.
As the scaling we
see in the previous section holds for generic product states, let
us choose the ground state of this set to be our initial state,
i.e. we begin with the state 
$\frac{1}{2}(\ket{00}+\ket{11})_1(\ket{01}+\ket{10})_2$ in
$(\sigma,\tau)$ language. For convenience, from now on we will
concatenate the basis states and express them as strings, for
example the state discussed above will now be written as
$\frac{1}{2}(\ket{0001}+\ket{0010}+\ket{1101}+\ket{1110})$.
Under the dynamics of just $\hat{H}_0+\hat{H}_V$, this state is left unchanged
as it is an eigenstate, and we shall call this as the initial
state $\ket{\psi_I}$. We will also assume that $V\gg J,\lambda$,
so that the sectors $+-,++,--,-+$ have energies of $-2V,0,0,2V$
respectively at leading order. Thus the sector under consideration
is well separated from the others energetically.

Now let us consider the effect of a simple transverse-field
perturbation $\hat{H}_1=\lambda\sum_i(\hat{\sigma}^x_i+\hat{\tau}^x_i)$ on our Hamiltonian dynamics.
At first order in $\lambda$, the eigenstates can be written as
\begin{equation}\label{eq:PT}
\ket{\ipr}=\ket{i}+\lambda\sum_{j\neq i}\frac{1}{E_i-E_j}\ket{j},
\end{equation}
where
the perturbed (original) basis is
denoted by $\ket{\ipr} (\ket{i})$. Note that the energy separation in
the sector structure as discussed above renders all $\ket{j}\neq\ket{i}$
contributing to the above expression to have a coefficient of $\lambda/V$,
as the perturbation is \textit{sector off-diagonal}, i.e, it has no
nonzero matrix elements within the sectors as defined by $\hat{H}_0$.
For ease of notation, let us express Eq.~\eqref{eq:PT} as $\ket{\ipr}=\sum_iM_{\ipr i}\ket{i}$. The relation between the perturbed and original basis
can also be similarly written using first order perturbation theory as 
$\ket{i}=\sum_\ipr M^{\prime}_{i\ipr}\ket{\ipr}$, and we will make use of both
these expression to understand the EE.

To understand the time evolution of our initial product state, let us
first express $e^{-i\hat{H}t}\ket{0}$, where $\hat{H}=\hat{H}_0+\hat{H}_1+\hat{H}_V$ and $\ket{0}$ now denotes the initial
state in the original basis, as
$\sum_\ipr e^{-iE^{\prime}_{\ipr}t}M^{\prime}_{0\ipr}\ket{\ipr}$. As the EE
is measured in the original basis by partitioning the lattice, we 
replace the dependence on the perturbed basis, and rewrite the time evolved
state as
$\sum_{\ipr,j}e^{-iE^{\prime}_{\ipr}t}M^{\prime}_{0\ipr}M_{\ipr j}\ket{j}$.
At intermediate times, we can consider the phase factor to be sufficiently
uncorrelated, and the dependence on $\lambda/V$ can be read directly from
the $M^{\prime}_{0\ipr}M_{\ipr j}$ terms, as no constructive or
destructive interference is to be expected. Of these, the largest coefficient
would be $M^{\prime}_{00^{\prime}}M_{0^{\prime}0}$, which is $\mathcal{O}(1)$.
To understand the other coefficients, let us reiterate a key property of
both $M$ and $M^{\prime}$, namely that if the matrix element corresponding
to $i,\ipr\,(\ipr,j)$ is nonzero, then it is $\mathcal{O}(1)$ if $i=\ipr\,(\ipr=j)$, and
$\lambda/V$ (to leading order) if not. This implies that for terms of order
$\lambda/V$, we only need to consider coefficients from matrix elements
$M^{\prime}_{00}M_{0j}$ and $M^{\prime}_{0\ipr}M_{\ipr\ipr}$.
As $M\,(M^{\prime})$ can only connect states within the starting sector to those outside,
we can write our resulting state after time evolution for an intermediate time
as $\ket{\psi_I}+(\lambda/V)\sum_\alpha c_\alpha\ket{\alpha}$, where $\alpha$ lists states outside
the sector of interest. 

\begin{figure}[t]
\centering
\includegraphics[width=\hsize]{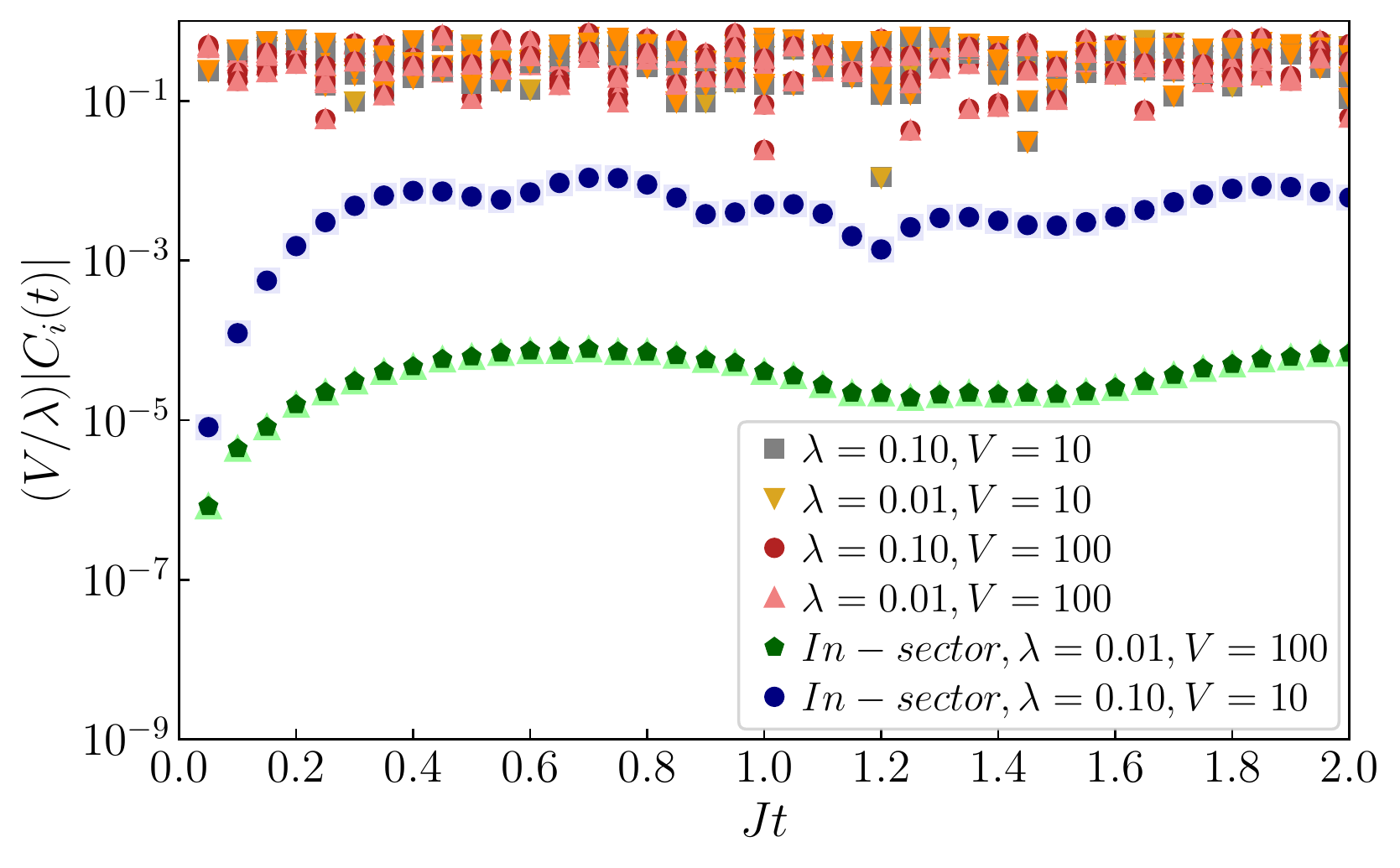}
\caption{Wave-function coefficients for relevant states as a
function of time for a $4$-rung
system. For the four nonequivalent
states out of sector, we see that a scaling with
$\lambda/V$ leads to
data collapse over a single decade, whereas for in-sector states 
(only those with
the largest coefficients shown), we clearly see that
the scaling is not sufficient, and that we require a higher power 
of $\lambda/V$ as expected.}
\label{figpert}
\end{figure}

The calculation of the half-chain EE of such a state will first require a reduced
density matrix. For the remainder of this Section, we shall use $c=\lambda/V$ for
ease of notation. The reduced density matrix for the state discussed above
can be expressed (to leading order) as

\begin{equation}
\hat{\rho}_A=\begin{pmatrix}
0.5 & ac & ac & 0.5 \\
ac & bc^2 & 0 & ac \\
ac & 0 & bc^2 & ac \\
0.5 & ac & ac & 0.5 \\
\end{pmatrix},
\end{equation}
where $a$ and $b$ are prefactors coming from the perturbative expansion, and the rows
correspond to the left rung configurations $00$, $01$, $10$, and $11$, respectively.
Diagonalizing this matrix reduces the von Neumann entropy $-\Tr[\hat{\rho}_A\log(\hat{\rho}_A)]$
to $-\sum_iE_i\log(E_i)$. The eigenvalues to $\mathcal{O}(c^2)$ are
$0,bc^2,a^{\prime}c^2,1+b^{\prime}c^2$, where $a^{\prime}$ and $b^{\prime}$ are
functions of $a$ and $b$. For the second and third eigenvalues, $E_i\log(E_i)$
reduces to the form $bc^2[2\log(c)+\log(b)]$, and the scaling with $c$ is
controlled approximately only by $c^2$, as the logarithm function is insensitive to variations in
$c$ for $c\ll1$. For the last eigenvalue, we can expand around unity, as $c\ll1$, and
this leads to $E_i\log(E_i)\approx \mathcal{O}(c^2)$. Thus we see that the overall scaling of
the von Neumann EE is controlled by $c^2$, i.e., $(\lambda/V)^2$.

As the above analysis is for a two-rung system, and carrying out the same argument for
a larger system size quickly becomes analytically intractable, we would like to
confirm that the assumption about the state at time $t$, which we have made above,
holds true. This would justify our expectation that a similar argument extends to
the thermodynamic limit. For ease of understanding, let us assume that we initialize
our state as the ground state in the sector $+-+-$. This is again a product state
of the symmetric eigenvector of $\sum_i\hat{\sigma}^x_i\hat{\tau}^x_i$. Explicitly, our initial
state is now 

\begin{align}
\ket{\psi_I}{=}\frac{1}{4}(\ket{00}{+}\ket{11})_1(\ket{01}{+}\ket{10})_2(\ket{00}{+}\ket{11})_3(\ket{01}{+}\ket{10})_4,
\end{align}
where the subscript denotes the rung number. Upon expanding the above expression, it
is easy to see that this is a uniform superposition over the 16 basis states making up
the sector $+-+-$. To ensure that other states within this sector do not participate in
the time evolution, we can plot the coefficients of the 16 basis states, and if they
stay equal throughout the time evolution for the period of interest, we can conclude
that the other in-sector states can be neglected. For the states that are connected at
first order in perturbation theory, careful observation of the structure of the Hamiltonian in state
space shows a number of states that have equivalent dynamic behavior. We
choose four nonequivalent states out of these: 
$\ket{01010001},\ket{10010001},\ket{00000001}$, and $\ket{00110001}$. We expect
to find that the coefficients of these states are $\propto(\lambda/V)^2$. The results
for the coefficients discussed above from our exact numerics are shown in
Fig.~\ref{figpert},
and indeed we find that (i) the coefficients of the 15 in-sector eigenstates of $H_0$ require a higher power of $V/\lambda$
to get a scaling collapse,
implying that in-sector states play no role at first order, and (ii) the 4 out-of-sector states
discussed above have a coefficient that scales with $(\lambda/V)$. Thus we find
that the assumptions of our perturbative analysis are correct, and that we can expect
the results to extend to larger system sizes.

\section{Conclusions}\label{sec:conc}

We have investigated a fine-tuned many-body spin system, which behaves as a system
of smaller noninteracting subsystems due to a classical degeneracy in inter-site
interactions, and carefully engineered local conserved quantities. While the former is
expected to occur for generic insulating crystals due to space group symmetries,
the latter severely restricts the form of quantum fluctuations available to the system.
Thus we have studied the effect of generic perturbations (of strength $\lambda$)
that destroy the local
conservation, and lead to a standard growth in entanglement with increasing evolution time,
as would be expected for a generic quantum many-body system.

To recover features of our disentangled dynamics, we impose a large energy separation
$V$ via a sector protection term,
between the sector (indexed by the values of the local conserved quantities) of interest
and the rest of the spectrum. This effectively isolates our sector dynamically, leading
to a suppression of the effect of the perturbation which couples various sectors.
Numerical investigations show that the entanglement developed is effectively controlled
by $\lambda/V$, where the entanglement entropy settles into a prethermal plateau $\propto\lambda^2/V^2$ up to timescales polynomial in $V$. We justify this behavior via a first-order perturbative analysis. Although our protective term is chosen to be specific to the sector we want to protect, we also see that there is an accidental protection of other sectors that have the
disentangled structure. This is specific to the form of the perturbation, which
is unable to mix sectors within the same energy plateau at first order and as higher-order contributions are suppressed by the energy scale $V$. However, we also see that for
different forms of the perturbation, which are able to transition between sectors in the same energy plateau, the entanglement reaches its saturation value in time $\propto$ system size.

The example that we have studied demonstrates the significance of this special fine-tuned point in parameter space, and the extent to which it can control the dynamics in a perturbative regime. This can easily be extended to larger systems and geometries in higher dimensions, as the local conserved quantity acts only on-site. Given that
spin systems are realized in generic insulating crystals, it is also possible to find anisotropic crystals of the XXZ type which would be proximate to the Hamiltonian discussed here.

Finally, let us point out that this mechanism can prove to be extremely useful for
quantum computing setups where one wishes to eliminate unwanted interactions between
qubits. Thus, future directions include designing gates for this rung architecture, and considering possible experimental platforms where we may be able to realize such physics.

\begin{acknowledgments}
A.S.~acknowledges funding by IFW Excellence Programme 2020 and Ulrike Nitzsche for technical assistance. J.C.H.~is grateful to Philipp Hauke and Haifeng Lang for work on related projects. J.C.H.~acknowledges funding within the QuantERA II Programme that has received funding from the European Union’s Horizon 2020 research and innovation programme under Grand Agreement No 101017733, support by the QuantERA grant DYNAMITE, by the Deutsche Forschungsgemeinschaft (DFG, German Research Foundation) under project number 499183856, funding by the Deutsche Forschungsgemeinschaft (DFG, German Research Foundation) under Germany's Excellence Strategy -- EXC-2111 -- 390814868, and funding from the European Research Council (ERC) under the European Union’s Horizon 2020 research and innovation programm (Grant Agreement no 948141) — ERC Starting Grant SimUcQuam.
\end{acknowledgments}

\bibliography{biblio}

\end{document}